\documentclass{PoS}

\usepackage{axodraw}

\usepackage[caption=false]{subfig}

\usepackage{dcolumn}
\usepackage{amsmath}
\newcommand{\lgm}{{\,\rm ln }}

\newcommand{\re}[1]{(\ref{#1})}

\newcommand{\ice}[1]{\relax}

\newcommand{\as}{a_s}

\newcommand{\als}{\alpha_s}
\newcommand{\beq}{\begin{equation}}

\newcommand{\ba}{\begin{array}}

\newcommand{\ea}{\end{array}}

\newcommand{\eeq}{\end{equation}}

\newcommand{\bea}{\begin{eqnarray}}

\newcommand{\eea}{\end{eqnarray}}
\newcommand{\g}{\gamma}

\newcommand{\al}{\alpha}

\newcommand{\EQN}{\label}
\newcommand{\ovl}{\overline}

\newcommand{\BreakI}{ \right. \nonumber \\ &{}& \left. }

\newcommand{\sbz}{  }

\title{$R(s)$  and Z decay in order $\alpha_s^4$: complete results }

\ShortTitle{R(s) and Z decay in order $\alpha_s^4$}

\ice{
\author{\speaker{Konstantin Chetyrkin}
         \thanks{WWWWWWWWWWw.}\\
        Institut fuer Theoretische Teilchenphysik\\
        E-mail: \email{chet@particle.physik.uni-karlsruhe.de}}

}

\author{P.~A.~Baikov,$^a$ \speaker{K.~G.~Chetyrkin},$^b$\thanks{
E-mail: {konstantin.chetyrkin@kit.edu}}
\, J.~H.~K\"uhn$^b$
and J.~Rittinger$^b$\\
\llap{$^a$} Skobeltsyn Institute of Nuclear Physics,
Lomonosov Moscow State University,
Moscow 119991, Russia \\
\llap{$^b$}Institut f\"ur Theoretische Teilchenphysik, Karlsruhe
  Institute of Technology (KIT), Wolfgang-Gaede-Stra\ss{}e 1, 726128 Karlsruhe, Germany}

\abstract{
We report on our calculation of the order $\alpha_s^4$ axial singlet
contributions for the decay rates of the $Z$-boson as well as the
vector singlet contribution  to the cross section for
electron-positron annihilation into hadrons. Together with recently
finished ${\cal O}(\alpha_s^4)$ calculations of the non-signlet
corrections \cite{Baikov:2008jh,PhysRevLett.104.132004}, the new
results directly lead us to the first {\em complete} ${\cal
O}(\alpha_s^4)$ predictions for the total hadronic decay rate of the
Z-boson and the ratio
$ R(s) =
{\sigma(e^+e^-\to {\rm hadrons})\over \sigma(e^+e^-\to \mu^+\mu^-)}\,
$.
}

\FullConference{ 10th International Symposium on Radiative Corrections (Applications of Quantum Field Theory to Phenomenology) - Radcor2011\\
September 26-30, 2011\\
Mamallapuram, India}

\begin{document}

\section{Introduction}

Inclusive quark production through a decay of a heavy virtual
photon,  Z boson or  $\tau$   is a process of importance for QCD as the theory
of strong interactions.  Perturbative QCD (pQCD)  provides
theoretically clean prediction for the process (see, e.g. \cite{Chetyrkin:1996ia,Davier:2005xq}). 

Combined with the
precise determination of the $Z$-boson decay rate into hadrons at LEP
\cite{Alcaraz:2007ri} this  has  led to one of the most precise
determinations of the strong coupling constant $\alpha_s(M_Z)$. An
alternative and also very precise determination of $\alpha_s(M_Z)$ as derived from 
$\alpha_s(M_\tau)$ 
has been recently obtained from the ${\cal O}(\alpha_s^4)$ prediction
\cite{Baikov:2008jh} for the ratio 
$R_{\tau} =
\frac{\Gamma(\tau\rightarrow{\rm hadrons})}{\Gamma(\tau\rightarrow l+\bar\nu_l+\nu_\tau)}
$
and the experimental determinations of $R_{\tau}$  by ALEPH, CLEO and OPAL 
collaborations (see, e.g. \cite{Davier:2005xq}).

Note that  while the ${\cal O}(\alpha_s^4)$ predictions for
$R_{\tau}$ are complete this is not the case for $R(s)$ and the Z-decay
rate.  The missing pieces are related to so-called singlet diagrams
(see Fig.~1 below).  Note that while the top quarks can not be produced in
Z-decays due to kinematical reasons, the (axial) singlet diagrams
containing internal quark loops are {\em not} power suppressed (unlike
similar loops for the vector singlet (and non-singlet) diagrams. This
remarkable  phenomenon first shows up at order $\alpha_s^2$
and   was first established and fully investigated in works \cite{Kniehl:1989bb,Kniehl:1989qu}. 
The full  account of 
singlet diagrams at order  $\alpha_s^3$  was performed in papers 
\cite{Gorishnii:1990vf,Surguladze:1990tg,Larin:1994va} 
(vector case) and in \cite{Chetyrkin:1993jm,Chetyrkin:1993ug,Larin:1993ju,Larin:1994va}
 (axial case).

In the present work we present the results of the calculations of the
order $\alpha_s^4$ axial singlet contributions for the decay rates of
the $Z$-boson as well as the vector singlet contribution to the cross
section for electron-positron annihilation into hadrons. Note that we
will not dwell on any phenomenological applications of our
calculations as they have been recently discussed in some detail in
\cite{Baikov:2012er}.

\section{Preliminaries}

The interaction of the Z boson to quarks is described
(in the lowest order approximation in the weak coupling constant)
by adding to  the QCD Lagrangian an extra term of the form
$M_Z\left(\frac{G_F}{2\sqrt 2}\right)^{1/2}Z^\alpha J^0_\alpha $,
with
$
J^0_\al
=
\sum_i \ovl{\psi}_i\g_\al(g^V_i -   g^A_i\g_5)\psi_i \
$
being the neutral  quark current.
As a result, the hadronic decay rate of the Z boson ($\Gamma^h_Z$)
including  all strong interaction corrections
may be viewed as an incoherent sum of  vector
($\Gamma^V_Z$) and axial ($\Gamma^A_Z$)  contributions.
By the optical theorem  both contributions can be conveniently related to 
the correlators of vector and axial vector quark currents. The general definition for the
latter reads:
\begin{align}
\Pi_{\mu\nu;i,j}^{V/A}(q)&= i\int e^{iqx}\langle0\vert~ T~j_{\mu,i}^{V/A}(x)j_{\nu,j}^{V/A}(0)~\vert0\rangle~\mathrm{d}x \nonumber
  \\
 &= g^{\mu\nu}q^2\Pi^{V/A}_{1;i,j}(-q^2)+q^\mu q^\nu\Pi^{V/A}_{2;i,j}(-q^2) \label{correlator}
\end{align}
with $j_{\mu,i}^{V} = \overline{\psi}_i~\gamma_\mu~\psi_i=V^i_\mu$ and
$j_{\mu,i}^{A} = \overline{\psi}_i~\gamma_\mu~\gamma_5\psi_i=A^i_\mu$.
The corresponding absorptive parts are  defined as follows:
\begin{align}
R^{V/A}_{i,j}(s)= 12\pi \Im \  \Pi^{V/A}_{1;i,j}(-s- i\,\varepsilon) 
{}. 
\label{RVA}
\end{align}
The $Z$ decay rate $\Gamma(Z\rightarrow\textnormal{hadrons})=\Gamma_0 \bigl(\, R^V(M_Z^2) + R^A(M_Z^2)\bigr)$, 
where $\Gamma_0=G_F M_Z^3/(24\pi\sqrt{2})$ and $R^{V/A}$  can be expressed in terms of $R^{V/A}_{i,j}$ defined
in eq.~(\ref{RVA}), namely
\begin{align}
 R^V=\sum_{i,j} g_i^V\,g_j^V\,R^V_{i,j}~,\qquad\qquad R^A=\sum_{i,j} g_i^A\,g_j^A\,R^A_{i,j}~.
\end{align}

Similarly, the inclusive cross-section reaction of the reaction $e^+e^-$ 
annihilation into hadrons through the photon is described by the current correlation function 
\begin{equation}
\label{Pi}
\Pi_{\mu\nu}(q)  =
\int {\rm d} x e^{iqx}
\langle 0|T[ \;
\;j_{\mu}^{\rm em}(x)j_{\nu}^{\rm em}(0)\;]|0 \rangle
= 
\displaystyle
(-g_{\mu\nu}q^2  + q_{\mu}q_{\nu} )\Pi^{EM}(-q^2)
{}\, ,
\end{equation}
with the hadronic EM current
\[
j^{\rm em}_{\mu}=\sum_{{f}} q_{{f}}
\overline{\psi}_{{f}} \gamma_{\mu} \psi_f
\ \ \
\mbox{and} 
\ \ \
R(s) = 12 \pi\,\Im \Pi^{EM}(-s- i\,\varepsilon) 
{},
\]
with 
$q_f$ being the EM charge of the quark $f$.
As a result, we arrive to the following representation for the ratio
R(s) valid in massless approximation (precise definitions of $R^{NS}$
and $R^{V,S}$ will be given below)
\beq
R(s) = \sum_{i,j} q_i\,q_j\,\,R^V_{i,j}(s)
= \Bigl(\sum_i q_i^2 \Bigr)\, R^{NS}(s) + \Bigl(\sum_i q_i\Bigr)^2 \, R^{VS}(s) 
{}
.
\eeq

\begin{figure}[t]
\centering
\subfloat[]{
\begin{picture}(70,40)(0,0)
  \Photon(0,20)(15,20){3}{2}
  \Photon(55,20)(70,20){3}{2}
  \ArrowArc(35,20)(20,0,180)
  \ArrowArc(35,20)(20,180,360)
  \Gluon(25,37)(25,3){3}{4}
  \Gluon(45,37)(45,3){-3}{4}
  \Text(5,25)[rb]{$Z$}
  \Text(65,25)[lb]{$Z$}
\end{picture} }
\hspace{3mm}
\subfloat[]{
\begin{picture}(70,40)(0,0)
  \Photon(0,20)(15,20){3}{2}
  \Photon(55,20)(70,20){3}{2}
  \ArrowArc(35,20)(20,0,180)
  \ArrowArc(35,20)(20,180,360)
  \Gluon(35,40)(35,28){-3}{1}
  \Gluon(35,0)(35,12){-3}{1}
  \ArrowArc(35,20)(8,0,360)
  \Text(5,25)[rb]{$Z$}
  \Text(65,25)[lb]{$Z$}
  \Text(45,20)[lc]{$t$}
\end{picture} }
\hspace{3mm}
\subfloat[]{
\begin{picture}(80,40)(0,0)
  \Photon(0,20)(15,20){3}{2}
  \Photon(65,20)(80,20){3}{2}
  \ArrowLine(15,20)(30,35)
  \Line(30,35)(30,05)
  \ArrowLine(30,05)(15,20)
  \ArrowLine(50,35)(65,20)
  \Line(50,05)(50,35)
  \ArrowLine(65,20)(50,05)
  \Gluon(30,20)(50,20){3}{2}
  \Gluon(30,35)(50,35){3}{2}
  \Gluon(30,5)(50,5){-3}{2}
  \Text(5,25)[rb]{$Z$}
  \Text(75,25)[lb]{$Z$}
\end{picture} }
\hspace{3mm}
\subfloat[]{
\begin{picture}(100,40)(0,0)
  \Photon(0,20)(15,20){3}{2}
  \Photon(85,20)(100,20){3}{2}
  \ArrowLine(15,20)(30,35)
  \Line(30,35)(30,05)
  \ArrowLine(30,05)(15,20)
  \ArrowLine(70,35)(85,20)
  \Line(70,05)(70,35)
  \ArrowLine(85,20)(70,05)
  \Gluon(30,20)(70,20){3}{4}
  \Gluon(30,5)(70,5){-3}{4}
  \Gluon(30,35)(42,35){3}{1}
  \Gluon(58,35)(70,35){3}{1}
  \ArrowArc(50,35)(8,90,450)
  \Text(5,25)[rb]{$Z$}
  \Text(95,25)[lb]{$Z$}
  \Text(58,42)[lb]{$t$}
\end{picture} }
\caption{ 
Examples of  non-singlet diagrams (a), (b), where the two $Z$ vertices
are connected by a fermion line, and of  singlet diagrams (c),(d),
where the diagram can be split by only cutting gluon lines. The
imaginary part of the non-singlet diagrams gives $R^{V/A,NS}$, while
the imaginary part of the singlet diagrams is denoted by $R^{V/A,S}$.} 
\label{propNSaS}
\end{figure}
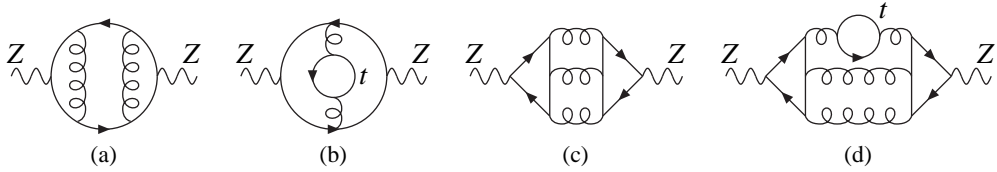
As the Z-boson is much heavier than all known quarks but the top one, it
is natural\footnote{ Mass corrections to both vector and axial vector
correlator due to other massive quarks are dominated by the bottom
quark and can be classified by orders in $m_b^2/M_Z^2$ and $\alpha_s$.
Up to ${\cal O}(\als^2 m_b^2/M_Z^2)$ and ${\cal O}(\als^2
m_b^4/M_Z^4)$ they can be found in
\cite{Chetyrkin:1996ia}, as well terms of  order 
$\als^2 m_b^2/M_Z^2$ (const + $\log\   m_b^2/M_Z^2$)
and  $\als^2 m_b^2/M_t^2$ (const + $\log\  m_b^2/M_Z^2$)
that arise  from  axial vector singlet contributions. Terms  of order 
$\als^3 m_b^4/M_Z^4$ and $\als^4 m_b^2/M_Z^2$ can be found in  
\cite{Chetyrkin:2000zk} and
\cite{Baikov:2004ku} respectively.
Corrections of order $\als^2 m_Z^2/m_t^2$ and  $\als^3 m_Z^2/m_t^2$ from 
singlet and non-singlet terms are known from \cite{Kniehl:1989qu,Kniehl:1989bb,Chetyrkin:1993tt} 
and \cite{Larin:1994va} respectively.
} to neglect  all power suppressed light quark mass corrections when  dealing with ${\cal O}(\alpha_s^4)$ contribution to 
$\Gamma^h_Z$. 
It is  customary to  split 
$R_{i,j}^{V/A}$ into two contributions as described in Fig.~\ref{propNSaS}
\begin{align}
 R_{i,j}^{V/A}(M_Z^2)=\delta_{ij}^{\ell}\,R^{NS}(M_Z^2)+R^{V/A,S}_{ij}(M_Z^2)~
\label{Rij}
{},
\end{align}
with the delta function $\delta_{ij}^{\ell} \equiv \delta_{ij}$ if both flavours
$i$ and $j$ are light and $\delta^{\ell}_{ij} =0$ if either $i$ or/and $j$  refer to the 
top quark.
In the non-singlet diagrams there is no top quark present in the
fermion loop connecting the two external currents, because these
diagrams have no physical cut and therefore have no imaginary part
contributing to $R^{NS}(s=M_Z^2)$. This, together with the assumed
masslessness of all quarks but top leads to the factorized form of the
non-singlet term in eq.~(\ref{Rij}). Note that the {\em internal} top
quark loops like in diagram (b) of Fig. 1 still contribute to
$R^{NS}(s=M_Z^2)$ if the strong coupling  is defined for 6 flavours. However, it is well known that such contributions
could be naturally described (up to power suppressed terms) by
transition from the full $n_f=6$ QCD to the effective massless $n_f=5$
one (see, e.g. \cite{Chetyrkin:1996ia} and references therein). In
fact, the same is true for the vector singlet term 
($\theta^h_{ij}$ below is defined as 1 if either i or/and j refer to the 
top quark and 0 in all other cases)
\[
 R^{V,S}_{ij}(M_Z) \equiv  (1-\theta^h_{ij})\, R^{V,S}(M_Z) + {\cal O}(M_Z^2/M^2_t)
{}.
\]
\ice{
(the power suppressed corrections in the above relation are known to 
order $\alpha^3$ from \cite{}).
}

The corresponding massless calculations of $R^{NS}$ in order
$\alpha_s^4$ have been recently finished \cite{Baikov:2008jh,PhysRevLett.104.132004}. 
In what follows we
concentrate on the singlet terms  $R^{V,S}$ and $R^{A,S}$.

\section{$\gamma_5$-treatment}

As is well-known the treatment of $\gamma_5$ within dimensional regularization is a non-trivial 
problem by itself (for an excellent review see  \cite{Jegerlehner:2000dz}). Following works 
\cite{Chetyrkin:1993jm,Chetyrkin:1993ug}
in all  our calculations we employ, in fact, two different definitions of $\gamma_5$.
First, for all non-singlet diagrams completely  anticommuting naive  $\gamma_5$ have been used.
Second, for singlet diagrams 
we employ essentially the 't Hooft-Veltman  definition  ~\cite{'tHooft:1972fi}
\beq
A^i_\alpha = \ovl{\psi}_i\g_\alpha\g_5 \psi_i \equiv
\frac{\xi^A_5(\as)}{6}\,  {\mathrm{i} } \, \epsilon _{\alpha\beta\nu\rho}
\ovl{\psi_i}\g_\beta\g_\nu\g_\rho \psi_i.
\EQN{axial.c.def}
{},
\eeq
where the current $\ovl{\psi_i}\g_\beta\g_\nu\g_\rho \psi_i$ is
assumed to be minimally renormalized. 

The finite normalization factor $\xi_5^A = 1-  4{\as}/{3}
+O(\as^2)$ on the rhs of  \re{axial.c.def}
is necessary  \cite{Trueman:1979en,Larin:1993tq}
for  the current  \re{axial.c.def} to obey the usual
(non-anomalous) Ward identities which in turn are crucial
in  renormalizing the Standard  Model.

In principle, one could (and even have to!)  use one and the
same definition  \re{axial.c.def} also for non-singlet
diagrams.  This would result to much  more complicated calculations due to
significantly longer traces encountered. Fortunately, it is not necessary
 because the factor
$\xi_5^A $ is {\em chosen} in such a  way to restore the anti-commutativity of the $\gamma_5$
(for a detailed discussion, see \cite{Larin:1993tq}).

\ice{which then can be
eliminated in the non-singlet diagrams. 

Note that  $R^{NS}$ is the same for the
vector and axial vector current (in the massless approximation),
}

\section{\ice{Calculation of the} Vector $\mathcal{O}(\alpha^4)$ singlet term  $R^{V,S}$}

From purely technical point of view the calculation  of the massless
five-loop diagrams  contributing to $\Pi^{V,S}_{ij}$ is not much different from
those contributing to $\Pi^{V,NS}$. Using  the  same  methods as described in 
\cite{Baikov:2008jh,PhysRevLett.104.132004}
we have obtained (below $a_s = {\alpha_s(\mu)}/{\pi}$ and $\mu$ is the renormalization scale in 
the $\overline{\mathrm{MS}}$  scheme)
\bea
R^{V,S}(s)  =  &{}& a_s^3 \,\Bigl(
\frac{55}{72} 
-\frac{5}{3}  \sbz \zeta_{3}
\Bigr)
\nonumber\\
&{+}& 
a_s^4\,\Biggl(
n_l\,\left[
-\frac{745}{432} 
+\frac{65}{24}  \sbz \zeta_{3}
+\frac{5}{6}  \,\zeta_3^2
-\frac{25}{12}  \sbz \zeta_{5}
-\frac{55}{144} \ln\frac{\mu^2}{s}
+\frac{5}{6}  \sbz \zeta_{3}\ln\frac{\mu^2}{s}
\right]
\nonumber\\
&{+}&
\frac{5795}{192} 
-\frac{8245}{144}  \sbz \zeta_{3}
-\frac{55}{4}  \,\zeta_3^2
+\frac{2825}{72}  \sbz \zeta_{5}
+\frac{605}{96} \ln\frac{\mu^2}{s}
-\frac{55}{4}  \sbz \zeta_{3}\ln\frac{\mu^2}{s}
\Biggr)
{}.
\label{Rsias4}
\eea

\section{\ice{Calculation of the} Axial vector $\mathcal{O}(\alpha^4)$ singlet term  $R^{A,S}_{i,j}$}

\ice{
Using the fact  that $g_i^A = 2\,I_3(i)$, with $I_3$ being the generator of the quark (weak) isospin,
the axial  part of the neutral current can  be
conveniently rewritten as a sum over quark weak
isodublets $(\psi_i,\psi_{i'})$
\beq
A^0_\alpha = \sum_{i=u,c,t}
\Delta^{i}_\alpha, \ \  \Delta^{i}_\alpha= A^i_\alpha - A^{i'}_\alpha
\EQN{2.1}
{}.
\eeq
}
Due to the obvious property\footnote{Obvious, thanks to the existence of
the unitary $SU(n_l)$ symmetry in the flavour  subspace of the first 
$1 \dots  n_l=5$ massless quarks.} $R^{A,S}_{i,j} = R^{A,S}_{i',j'}$ if
all 4 indexes refer to the massless quarks and the fact that $g_A^u + g_A^d = g_A^c + g_A^s = 0$,
 we can write the axial  singlet part of the Z decay rate as
follows:\footnote{Note that separate terms on the rhs of \re{R_AS_decomp2} are  
{\em{not scale-invariant}}, while their sum is \cite{Chetyrkin:1993jm,Chetyrkin:1993ug}.}
\beq
R^{A,S} = R^{A,S}_{tt} - 2\, R^{A,S}_{tb} + R^{A,S}_{bb} 
\label{R_AS_decomp2}
{}.
\eeq
All diagrams contributing to the first  two terms of  \re{R_AS_decomp2} contain at least
one top quark loop. The third term receives  contributions  by both the completely massless diagrams
and those with top quark loop (the latter start  from order  $\alpha_s^3$, an example is given by Fig. 1 (d)).

As $M_Z \ll 2\, M_t$, one can use the effective theory methods to
compute top-mass-dependent diagrams as a series in the ratio
$\frac{M_Z^2}{4 M_t^2}$. The procedure was elaborated long ago and
successfully employed (see works  \cite{Chetyrkin:1993hk,Chetyrkin:1993jm,Chetyrkin:1993ug}) 
to get all ingredients of eq.~\re{R_AS_decomp2}
at order $\alpha_s^3$  at leading order in $1/M_t$ expansion (still keeping {\em all}
power non-suppressed terms, including those 
which depends on $\ln (\mu^2/M_t^2)$). From purely technical point of view 
the evaluation at order $\alpha_s^4$ involves 
absorptive parts of five-loop diagrams with massless propagators and,
in addition,  absorptive parts of four-loop diagrams
combined with one-loop massive tadpoles, etc. down to one-loop
massless diagrams together with four-loop massive tadpoles. The latter
have been computed with the help  of the  Laporta's algorithm
\cite{Laporta:2001dd} implemented in Crusher \cite{crusher}. 
The  massive tadpoles with number of loops less or equal  three have been  independently recalculated with
the help of the FORM program MATAD \cite{Steinhauser:2000ry}.
Our results for  $R^{A,S}_{tt}$,   $R^{A,S}_{tb}$,  $R^{A,S}_{bb}$ and   $R^{A,S}$ read
\beq
{R^{A,S}_{tt} } 
=
(a_s^{5})^4
\left[
\frac{15}{64} 
-\frac{15}{8} \ell_{\mu t}
+\frac{15}{4}  \ell^ 2_{\mu t}
\right]
{},
\label{Rtt}
\eeq
\begin{eqnarray}
\lefteqn{R^{A,S}_{tb} =   
 \,(a_s^{5})^2
\left[
\frac{3}{8} 
-\frac{3}{2} \ell_{\mu t}
\right]
{+} \,(a_s^{5})^3
\left[
-\frac{3869}{288} 
+\frac{55}{8}  \sbz \zeta_{3}
-\frac{45}{8} \ell_{\mu t}
-\frac{25}{8}  \ell^ 2_{\mu t}
\right]
}
\nonumber\\
&{+}& \,(a_s^{5})^4
\left[
-\frac{370478273}{14515200} 
-  \sbz \zeta_{2}
+\frac{1309601}{16800}  \sbz \zeta_{3}
-\frac{4225817}{34560}  \sbz \zeta_{4}
-\frac{10453}{288}  \sbz \zeta_{5}
\BreakI
\phantom{+ \,(a_s^{5})^4}
-2  \sbz \zeta_{2} \,\ln (2)
-\frac{89}{48}  \sbz \zeta_{4} \,\ln (2)
-\frac{5861}{1080}  \sbz \zeta_{2} \,(\ln (2))^2
+\frac{2}{9}  \sbz \zeta_{2} \,(\ln (2))^3
+\frac{5861}{6480}  \,(\ln (2))^4
\BreakI
\phantom{+ \,(a_s^{5})^4}
-\frac{1}{45}  \,(\ln (2))^5
+\frac{5861}{270}  \,a_{4}
+\frac{8}{3}  \,a_{5}
-\frac{37}{32}  \ell_{\mu Z}\,
-\frac{47015}{576} \ell_{\mu t}
\BreakI
\phantom{+ \,(a_s^{5})^4}
+\frac{709}{8}  \sbz \zeta_{3}\ell_{\mu t}
+\frac{37}{8}  \ell_{\mu Z}\,\ell_{\mu t}
-\frac{363}{16}  \ell^ 2_{\mu t}
-\frac{193}{32}  \ell^ 3_{\mu t}
\right]
{},
\label{Rtb}
\end{eqnarray}
\begin{eqnarray}
\lefteqn{R^{A,S}_{bb} =  
 (a_s^{5})^2
\left[
-\frac{17}{2} 
-3  \ell_{\mu Z}\,
\right]
}
\nonumber\\
&{+}& \,(a_s^{5})^3
\left[
-\frac{4673}{48} 
+\frac{23}{2}  \sbz \zeta_{2}
+\frac{67}{4}  \sbz \zeta_{3}
-\frac{373}{8}  \ell_{\mu Z}\,
-\frac{23}{4}  \ell^2_{\mu Z}\, 
\BreakI
\phantom{+ \,(a_s^{5})^3}
-\frac{1}{12} \ell_{\mu t}
-\frac{1}{2}  \ell^ 2_{\mu t}
\right]
\nonumber\\
&{+}& \,(a_s^{5})^4
\left[
-\frac{79017683}{82944} 
+\frac{8747}{32}  \sbz \zeta_{2}
+\frac{54179}{128}  \sbz \zeta_{3}
+\frac{1481}{128}  \sbz \zeta_{4}
-\frac{6455}{96}  \sbz \zeta_{5}
\BreakI
\phantom{+ \,(a_s^{5})^4}
-  \sbz \zeta_{2} \,(\ln (2))^2
+\frac{1}{6}  \,(\ln (2))^4
+4  \,a_{4}
-\frac{174767}{288}  \ell_{\mu Z}\,
+\frac{529}{8}  \sbz \zeta_{2} \ell_{\mu Z}\,
\BreakI
\phantom{+ \,(a_s^{5})^4}
+\frac{1519}{8}  \sbz \zeta_{3} \ell_{\mu Z}\,
-\frac{8747}{64}  \ell^2_{\mu Z}\, 
-\frac{529}{48}  \ell^3_{\mu Z}\, 
-\frac{1975}{288} \ell_{\mu t}
+\frac{37}{8}  \sbz \zeta_{3}\ell_{\mu t}
\BreakI
\phantom{+ \,(a_s^{5})^4}
-\frac{247}{48}  \ell^ 2_{\mu t}
-\frac{25}{24}  \ell^ 3_{\mu t}
\right]
{},
\label{Rbb}
\end{eqnarray}
\begin{eqnarray}
\lefteqn{R^{A,S} =  
 \,(a_s^{5})^2
\left[
-\frac{37}{4} 
-3  \ell_{\mu Z}\,
+ 3  \ell_{\mu t}
\right]
\nonumber
}
\\
&{+}& \,(a_s^{5})^3
\left[
-\frac{5075}{72} 
+\frac{23}{2} +  \sbz \zeta_{2}
+3   \sbz \zeta_{3}
-\frac{373}{8}  \ell_{\mu Z}\,
-\frac{23}{4}  \ell^2_{\mu Z}\, 
\BreakI
\phantom{+ \,(a_s^{5})^3}
+\frac{67}{6} \ell_{\mu t}
+\frac{23}{4}  \ell^ 2_{\mu t}
\right]
\nonumber\\
&{+}& \,(a_s^{5})^4
\left[
-\frac{13083735979}{14515200} 
+\frac{8811}{32}  \sbz \zeta_{2}
+\frac{17967167}{67200}  \sbz \zeta_{3}
+\frac{553219}{2160}  \sbz \zeta_{4}
+\frac{1541}{288}  \sbz \zeta_{5}
\BreakI
\phantom{+ \,(a_s^{5})^4}
+4  \sbz \zeta_{2} \,\ln (2)
+\frac{89}{24}  \sbz \zeta_{4} \,\ln (2)
+\frac{5321}{540}  \sbz \zeta_{2} \,(\ln (2))^2
-\frac{4}{9}  \sbz \zeta_{2} \,(\ln (2))^3
-\frac{5321}{3240}  \,(\ln (2))^4
\BreakI
\phantom{+ \,(a_s^{5})^4}
+\frac{2}{45}  \,(\ln (2))^5
-\frac{5321}{135}  \,a_{4}
-\frac{16}{3}  \,a_{5}
-\frac{174101}{288}  \ell_{\mu Z}\,
+\frac{529}{8}  \sbz \zeta_{2} \ell_{\mu Z}\,
\BreakI
\phantom{+ \,(a_s^{5})^4}
+\frac{1519}{8}  \sbz \zeta_{3} \ell_{\mu Z}\,
-\frac{8747}{64}  \ell^2_{\mu Z}\, 
-\frac{529}{48}  \ell^3_{\mu Z}\, 
+\frac{11125}{72} \ell_{\mu t}
-\frac{1381}{8}  \sbz \zeta_{3}\ell_{\mu t}
\BreakI
\phantom{+ \,(a_s^{5})^4}
-\frac{37}{4}  \ell_{\mu Z}\,\ell_{\mu t}
+\frac{2111}{48}  \ell^ 2_{\mu t}
+\frac{529}{48}  \ell^ 3_{\mu t}
\right]
{}.
\label{Rax}
\end{eqnarray}
Here $a^5_s = {\alpha_s(\mu)}/{\pi}$ in the effective (topless) $n_f=5$ QCD, $\ell_{\mu Z} = \ln
\frac{\mu^2}{M_Z^2}$, $\ell_{\mu t} = \ln \frac{\mu^2}{M_t^2}$, and
$M_t$ is the pole top quark mass.
In addition,  $\zeta_n = \zeta(n)$ is Riemann's zeta function and 
$a_n = {\rm Li}_n(1/2) = \sum_{i=1}^{\infty} 1/(2^i i ^n)$.


Finally, setting $ \mu=M_Z $, we arrive at the following numerical form of \re{Rax}
\begin{eqnarray}
\lefteqn{R^{A,S} =  
 \,(a_s^{5})^2
\left[
-9.25 
+3.  \lgm\frac{M_Z^2}{M_t}
\right]
}
\nonumber\\
&{+}& \,(a_s^{5})^3
\left[
-47.9632 
+11.1667  \lgm\frac{M_Z^2}{M_t}
+5.75  \lgm^2\frac{M_Z^2}{M_t}
\right]
\nonumber\\
&{+}& \,(a_s^{5})^4
\left[
147.093 
-52.9912  \lgm\frac{M_Z^2}{M_t}
+43.9792  \lgm^2\frac{M_Z^2}{M_t}
+11.0208  \lgm^3\frac{M_Z^2}{M_t}
\right]
{}.
\label{RaxN}
\end{eqnarray}

\section{Conclusion}

All our calculations have been
performed on a SGI ALTIX 24-node IB-interconnected cluster of 8-cores
Xeon computers and on the HP XC4000 supercomputer of the federal state
Baden-W\"urttemberg
using  parallel  MPI-based \cite{Tentyukov:2004hz} as well as thread-based \cite{Tentyukov:2007mu} versions  of FORM
\cite{Vermaseren:2000nd}.  
For evaluation of color factors we have used the FORM program { COLOR}
\cite{vanRitbergen:1998pn}. The diagrams have been generated with QGRAF \cite{Nogueira:1991ex}.
This work was supported by the Deutsche Forschungsgemeinschaft in the
Sonderforschungsbereich/Transregio SFB/TR-9 ``Computational Particle
Physics'' and  by RFBR grants   11-02-01196 and  10-02-00525.

We thank P. Marquard for  his friendly help with the package Crusher.

\end{document}